\documentclass[bibyear]{aa} 

\usepackage{graphicx}
\usepackage{txfonts}
\usepackage{longtable}
\begin{document}

   \title{The nature of the methanol maser ring G23.657$-$00.127}
    \subtitle{II. Expansion of the maser structure}

   \author{A. Bartkiewicz
          \inst{1},
          A. Sanna
          \inst{2,3},
          M. Szymczak
          \inst{1},
          L. Moscadelli
          \inst{4},
          H.J. van Langevelde
          \inst{5,}\inst{6}
          \and
          P. Wolak\inst{1}
          }

   \institute{Institute of Astronomy, Faculty of Physics, Astronomy and Informatics, Nicolaus Copernicus University, Grudziadzka 5, 87-100 Torun, Poland, \email{annan@astro.umk.pl}
         \and
         INAF, Istituto di Radioastronomia \& Italian ALMA Regional Centre, Via P. Gobetti 101, 40129, Bologna, Italy
         \and
         Max-Planck-Institut für Radioastronomie, Auf dem Hügel 69, 53121, Bonn, Germany
         \and
         INAF, Osservatorio Astrofisico di Arcetri, Largo E. Fermi 5, 50125, Firenze, Italy
         \and
             Joint Institute for VLBI ERIC (JIVE), Oude Hoogeveensedijk 4, 7991 PD Dwingeloo, The
Netherlands
         \and
             Sterrewacht Leiden, Leiden University, Postbus 9513, 2300 RA Leiden, The Netherlands
             }

   \date{Received 23 January 2020; accepted 20 March 2020}

 
  \abstract
{Ring-like distributions of the 6.7~GHz methanol maser spots at milliarcsecond scales represent a family of molecular structures of unknown origin associated with high-mass young stellar objects (HMYSOs).}
{We aim to study G23.657$-$00.127, which has a nearly circular ring of the 6.7~GHz methanol masers, and is the most suitable target to test hypotheses on the origin of the maser rings.} 
{The European Very Long Baseline Interferometry Network (EVN) was used at three epochs spanning 10.3\,yr 
to derive the spatio-kinematical structure of the 6.7\,GHz methanol maser emission in the target.}
{The maser cloudlets, lying in a nearly symmetric ring,
expand mainly in the radial direction with a mean velocity of 3.2~km~s$^{-1}$.
There is an indication that the radial component of the velocity increases with cloudlet's distance from the ring centre.
The tangential component does not show any clear evidence for rotation of the cloudlets or any relationship with distance from the ring centre. The blue-shifted masers may hint at an anticlockwise rotation of cloudlets in the southern part of the ring. The nearly circular structure of the ring clearly persisted for more than 10~yr. Interferometric data demonstrated that about one quarter of cloudlets show significant variability in their brightness, although the overall spectrum was non-variable in single-dish studies.
}
{Taking into account the three-dimensional motion of the maser cloudlets and their spatial distribution along a small ring, we speculate about two possible scenarios
 where the methanol masers trace either a spherical outflow arising from an (almost) edge-on disc, or a wide angle wind at the base of a protostellar jet. The latter is associated with near- and mid-infrared emission detected towards the ring. High angular resolution images of complementary (thermal) tracers are needed to interpret  the environment of methanol masers.
}
   \keywords{masers -- stars: massive -- instrumentation: interferometers -- stars: formation -- astrometry}

\titlerunning{Expansion of the G23.657$-$00.127 methanol maser ring}
\authorrunning{A. Bartkiewicz et al.}

   \maketitle
%

\section{Introduction}
Methanol masers at the 6.7~GHz transition are one of the signposts of high-mass star formation regions (e.g. Menten~\cite{m91}, Breen et al.~\cite{b13}), but it is not yet clear
what structures or processes in the protostellar environment they actually trace. For instance, interferometric studies of the masers revealed the kinematics to be
related to slow motions of radial expansion and rotation around a disc axis in G23.01$-$0.41 (Sanna et al.~\cite{s10b}), but dominant infall motions prevailed in
Cep~A$-$HW2 (Sugiyama et al.~\cite{su14}, Sanna et al.~\cite{s17}) and AFGL~5142~MM$-$1 (Goddi et al.~\cite{g11}). These results support two scenarios
where the maser emission comes from the intermediate region between the outflow and the rotating envelope in one case, and from the accretion disc in the other. 

A suite of maser morphologies has been observed in several high-mass young stellar objects (HMYSOs, e.g. Bartkiewicz et al.~\cite{b09}, \cite{b16}, Fujisawa
et al.~\cite{f14}, Pandian et al.~\cite{p11}), but G23.657$-$00.127 is particularly unique (Bartkiewicz et al.~\cite{b05}). The source was discovered in an unbiased  
survey of the 6.7~GHz methanol maser line (Szymczak et al.~\cite{sz02}). It displays a complex and relatively faint spectrum that is broader than 11~km~s$^{-1}$. 
Imaging with the European Very Long Baseline Interferometry Network\footnote{The European VLBI Network is a joint facility of independent European, African, Asian, and North-American radio astronomy institutes.
Scientific results from data presented in this publication are derived from the following EVN project codes: EN003, EB052.} (EVN) in 2004 revealed that the maser spots are
distributed in a nearly circular ring of $\sim$127~mas radius and a width of  $\sim$29~mas (Bartkiewicz et al.~\cite{b05}), which correspond to 405 and 95\,au,
respectively, for the trigonometric parallax distance of $3.19^{+0.46}_{-0.35}$\,kpc (Bartkiewicz et al.~\cite{b08}). Such a morphology can be interpreted as  the
methanol masers arising in a spherical bubble or in a disc seen nearly face-on or in an outflow oriented exactly towards the observer (Bartkiewicz et al.~\cite{b05}).
Gemini observations with a resolution of $\sim$150\,mas revealed that the peak of the 2.12\,$\mu$m emission nearly coincides with the centre of methanol ring, and
that the source morphology is fan shaped, suggesting scattered and/or reflected emission off the walls of the outflow cavity (De Buizer et al. ~\cite{d12}). The
bolometric luminosity of the source derived by Mottram et al.~(\cite{mot11a})  rescaled to the distance of 3.19\,kpc is 6$\times10^3$L$_{\odot}$ and is fully
consistent with De Buizer et al.'s (\cite{d12}) estimate implying a $\sim$10~M$_{\odot}$ central star of spectral type B1 (Mottram et al.~\cite{mot11b}, their Table~1).
Observations of $^{13}$CO (1$-$0) and NH$_3$ lines indicate a systemic velocity of 80.5~km~s$^{-1}$ (Urquhart et al.~\cite{u08}, \cite{u11}).

In this paper we report the results of EVN methanol maser observations of the target in order to determine the spatio-kinematical distribution of the masing gas.
Preliminary results, considering only two-epoch data, were presented in Bartkiewicz et al.~(\cite{b14}, \cite{b18}).

\section{Observations and data analysis}
\label{sec:obs}
The target was observed in the 6668.5192\,MHz methanol line at three epochs using the EVN (Table \ref{table1}). The phase-referencing technique was applied, using J1825$-$0737 as a reference. The bandwidth was set to 2~MHz and divided into 1024 channels, yielding a channel separation of 0.09~km~s$^{-1}$. In order to increase the signal-to-noise ratio (S/N) for the phase-reference source, dual circular polarisations were observed with eight 2\,MHz baseband converters during the second and third epochs.

The data were analysed with standard procedures for spectral line observations from the Astronomical Image Processing System (AIPS). The absolute accuracy of single spot positions was estimated to be about 2\,mas in RA (i.e. 0.1~ms) and 5\,mas in Dec (Bartkiewicz et al.~\cite{b09}). The absolute coordinates of the brightest spots with their uncertainties in each epoch are listed in Table~\ref{table1}. As we are interested in the detailed internal motions of masers, we instead relied on the relative motions. Therefore, 
FRING was used on the spectral channel with the most stable and brightest maser emission, the feature that is at the LSR velocity of ca.~82.5~km~s$^{-1}$. This relates all maser positions to this feature at every epoch and has the additional advantage that it excludes motions due to the annual parallax and Galactic rotation. Moreover, as the maser is bright enough, it also provides the highest possible signal-to-noise ratio (S/N). The accuracy of relative positions of the maser spots depends on the S/N ratio of the spots and the beam size and were better than 0.1\,mas at each epoch. 

In order to study the proper motions, we applied the following procedure, which was previously used successfully for different masers with diverse morphologies (e.g. Goddi et al.~\cite{g11}, Moscadelli et al.~\cite{m06}). The position, intensity, and radial velocity of all the maser spots in the source were determined for the three epochs using two-dimensional Gaussian fits. Groups of maser spots with S/N$>$10 were identified and called cloudlets 
if the
emission occurred in at least two contiguous spectral channels and coincided in position within half the synthesized beam (e.g. Sanna et al.~\cite{s17}).   
Persistent maser cloudlets were identified over the epochs when they showed reliable linear motions and a bright and not highly variable spectral feature. 
Motions were derived via linear fits of cloudlet displacements over the three epochs. Assuming that the maser motion is on average symmetric with respect to the HMYSO, the average proper motion of all persistent cloudlets was subtracted from the proper motion of each cloudlet. This way, we define a centre of motion approximating the HMYSO's rest frame, with respect to which we calculated the proper motions. This method is equivalent to calculating the geometric mean of position at single epochs for all maser cloudlets that persisted over three epochs and referring the proper motions to this point. 

Since the G23.657$-$00.127 shows widespread, almost circular symmetric methanol masers, we also made
use of fitted ellipses to the maser distribution in each epoch and derive the proper motions relatively to the centre of the ellipse. The ellipses were fitted to persistent (flux-weighted) cloudlets using the code by Fitzgibbon et al.~(\cite{f99}). Next, we aligned the centres of best fitted ellipses to the second and third epoch data with
respect to the first epoch and constructed the averaged proper motion
vectors, meaning the displacements between spots at two different epochs within a cloudlet were summed and divided by the number of spot pairs for each cloudlet. In this approach, we assumed that the centre of the ellipse coincided with the HMYSO position, and we removed any bulk motion of the ring in the plane of the sky to obtain the proper motions of masers relative to the HMYSO.

\begin{table*}
\centering
\caption{Observing parameters (Columns 1--4) and the absolute coordinates of the brightest spots with their uncertainties as measured in each epoch (Columns 5 \& 6).}
\label{table1}
\begin{tabular}{l@{\hspace{2pt}}lllcc@{}}
\hline
Exp. & Epoch(Date) & Beam &                 1$\sigma$ noise & RA (J2000) & Dec (J2000)\\
code &      & (mas$\times$mas); PA(\degr)&  (mJy b$^{-1}$) & (h:m:s) & ($^{\rm o}$: ': ")                    \\
\hline
EN003$^1$  & E1(2004 Nov 11) & 16$\times$5.5; $-$1\degr & 10 & 18:34:51.5648$\pm$0.0001 & $-$08:18:21.305$\pm$0.004 \\  
EB052  & E2(2013 Mar 2) & 9.0$\times$4.0; $-$28\degr & 4 & 18:34:51.5642$\pm$0.0001 & $-$08:18:21.325$\pm$0.004\\
EB052  & E3(2015 Mar 15) & 10.3$\times$3.6; $-$36\degr & 4 & 18:34:51.5641$\pm$0.0001 & $-$08:18:21.332$\pm$0.004 \\
\hline
\end{tabular}
\tablefoot{$^1$ taken from  Bartkiewicz et al. (\cite{b09})}
\end{table*}

\section{Results}

\begin{table*}
\centering
\caption{Parameters of flux-weighted ellipses fitted to the maser cloudlets with their uncertainties (1~$\sigma$) that appeared in all three epochs. Fits are presented in Fig.~\ref{three}.}
\label{ellipses}
\begin{tabular}{@{}clll@{}}
\hline
\multicolumn{1}{c}{Epoch} & \multicolumn{2}{c}{Centre$^a$} & \multicolumn{1}{c}{Semi-axes; PA$^b$}\\
 & \multicolumn{1}{c}{$\Delta$RA} & \multicolumn{1}{c}{$\Delta$Dec} &  \\
 & \multicolumn{1}{c}{(mas)} & \multicolumn{1}{c}{(mas)} & \multicolumn{1}{c}{(mas$\times$mas; \degr)} \\
\hline
 E1 & $-$66.058$\pm$0.025 & $-$96.411$\pm$0.012 & 134.89$\pm$0.022$\times$123.94$\pm$0.03; $-$20$\pm$0.1 \\
 E2 & $-$66.379$\pm$0.025 & $-$96.246$\pm$0.012 & 136.01$\pm$0.023$\times$124.49$\pm$0.023; $-$19$\pm$0.2 \\
 E3 & $-$66.824$\pm$0.024 & $-$97.350$\pm$0.012 & 136.06$\pm$0.012$\times$125.27$\pm$0.03; $-$19$\pm$0.1 \\
\hline
 \end{tabular}
\tablefoot{$^a$ The relative coordinates to the brightest spot position in the first epoch as given in Table~\ref{table1}. $^b$ The position
angle of the major axis (north to east).}
 \end{table*}

\begin{figure*}
\centering
\includegraphics[scale=0.51]{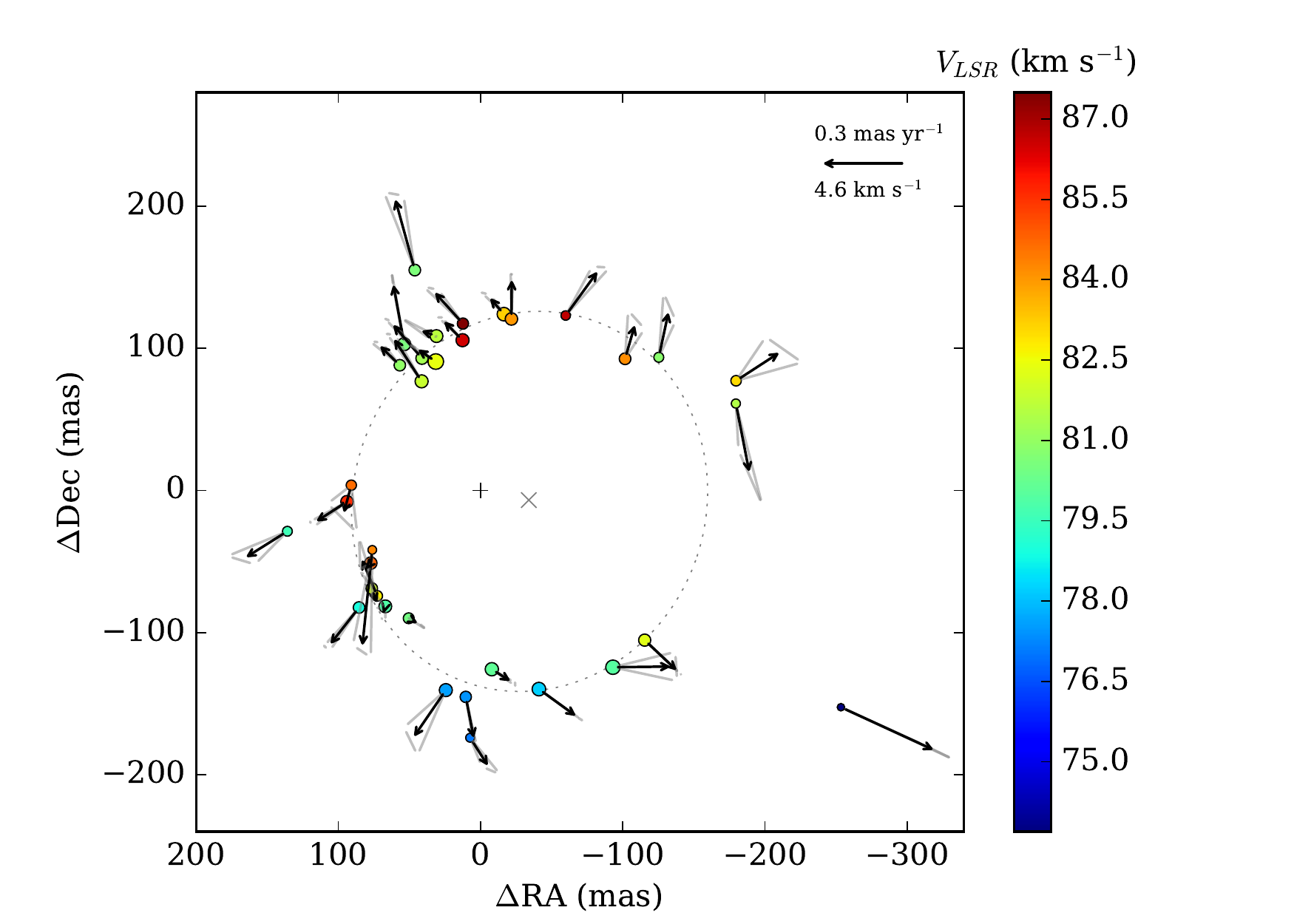}
\includegraphics[scale=0.51]{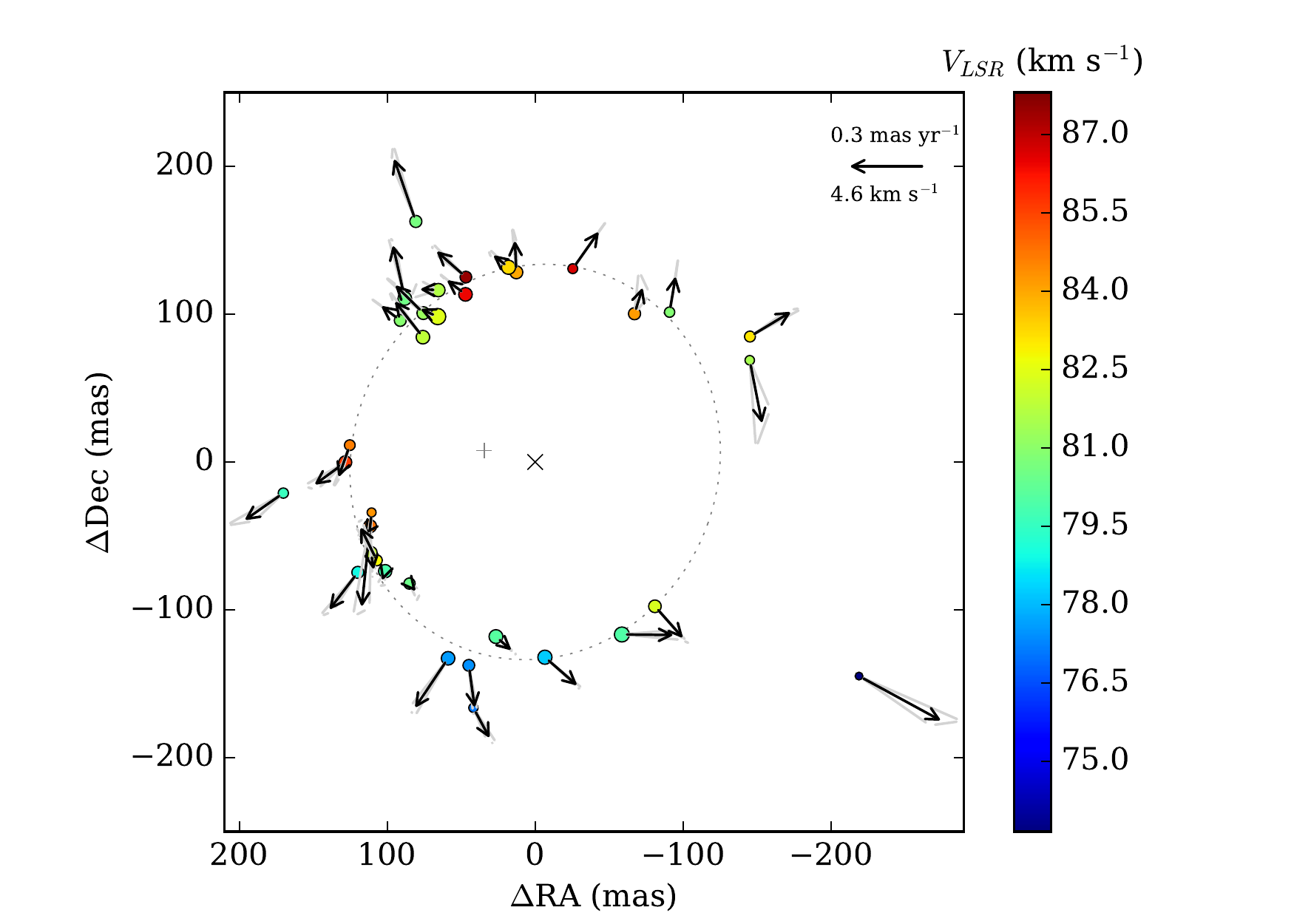}
\caption{{\bf Left:} Proper motions of 6.7~GHz methanol maser cloudlets in G23.657$-$00.127 as measured relative to the centre of motion. The centre of motion, defined in Sect.~\ref{sec:obs}, is marked by the plus sign. Its relative coordinates to the brightest spot in E1, as given in Table~\ref{table1}, are ($-$31.6145~mas, $-$88.7120~mas).
The cloudlets are marked by filled circles, their sizes are proportional to the peak brightness at the first epoch. Their colours correspond to the local standard velocity scale presented in the vertical wedge. The black arrows indicate the best fits of the relative proper-motion for the three epoch data with the uncertainties marked by the grey triangles. The dotted ellipse traces the best flux-weighted fit to all cloudlets (except the SW blue-shifted one) as detected at E1. The centre of the ellipse is marked by the cross (Table~\ref{ellipses}). {\bf Right:} Proper motions estimated via ellipse fittings and aligning their centres. The (0,0) point, marked by the plus sign, corresponds to the centre of the best fitted ellipse to E1 (Table~\ref{ellipses}). The black arrows represent the averaged proper motion vectors, defined as in Sect.~\ref{sec:obs}, between E1--E2 and E1--E3 data.}
\label{g23657cm}  
\end{figure*}

We detected more than 300 maser spots in the velocity range from 72.5 to 87.9~km~s$^{-1}$ at each epoch. The maps of the maser distribution shown in
Fig.\,\ref{three} indicate that the overall structure of the ring is persistent.

Using the first procedure described in Sect.~\ref{sec:obs},  
the positional offsets of the maser cloudlets at each observing epoch were calculated, and a linear fit of the
relative proper motion was done for each persistent cloudlet (Fig.~\ref{linearmotions}). The resulting proper motions of these cloudlets are displayed in Fig.~\ref{g23657cm}. The magnitudes of the proper motions of the maser cloudlets range from 0.04 to 0.36~mas~yr$^{-1}$, which correspond to a velocity of 0.63 to 5.4~km~s$^{-1}$ for the distance of 3.19~kpc. 

Consistent results of proper motions were obtained by applying the method of ellipse fitting, when we excluded the blue-shifted cloudlet to the SW. The major axis of the best fitted ellipses increased by about 2.34~mas, corresponding to 7.5~au between epochs E1 and E3. The parameters of the best fitted, flux-weighted ellipses, together with their uncertainties calculated using the bootstrap 
method for 20000 iterations (Press et al.~\cite{p07}), are listed in Table~\ref{ellipses}. The mean proper motions between epochs E1--E2 and E1--E3 are presented in Fig.~\ref{g23657cm}. The centre of the ellipse fitted to E1 data is only 35~mas shifted from the centre of motion. Since the first procedure, using the centre of motion, is independent from the maser morphology and takes into consideration all persistent cloudlets, we continue with these results.

\begin{figure*}
\centering
\includegraphics[scale=0.6, trim={0.8cm 0 0.5cm 0},clip]{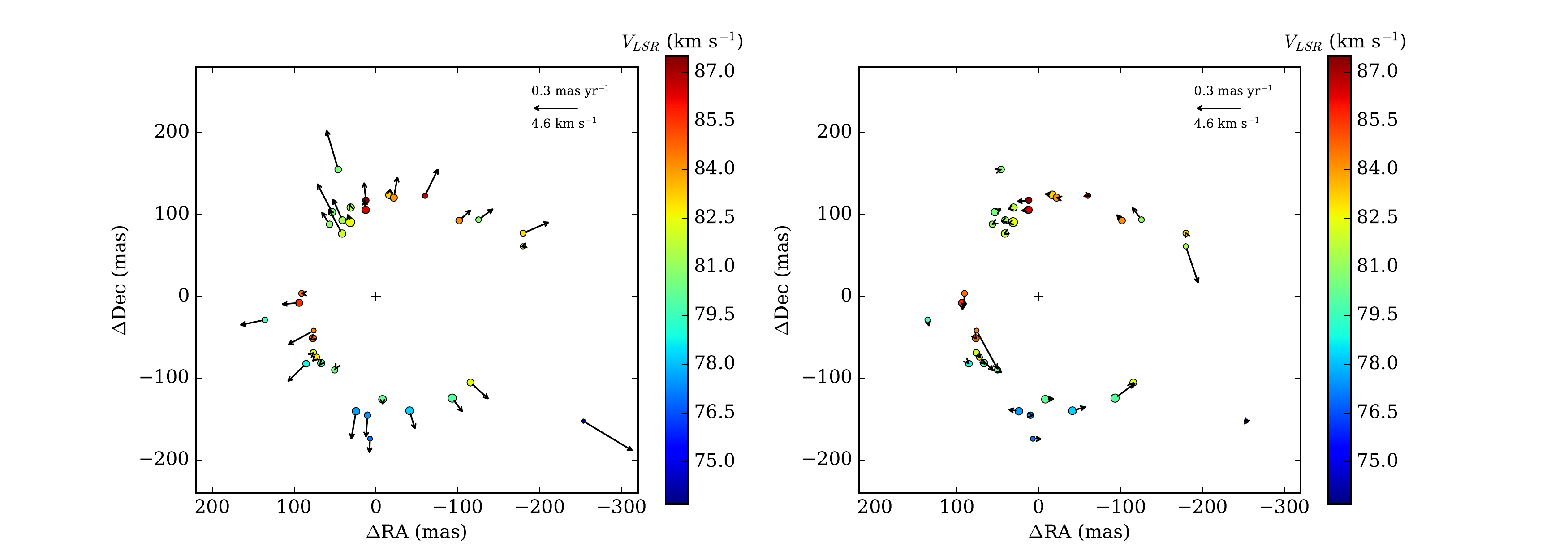}
\includegraphics[scale=0.6, trim={0.8cm 0 0.5cm 0},clip]{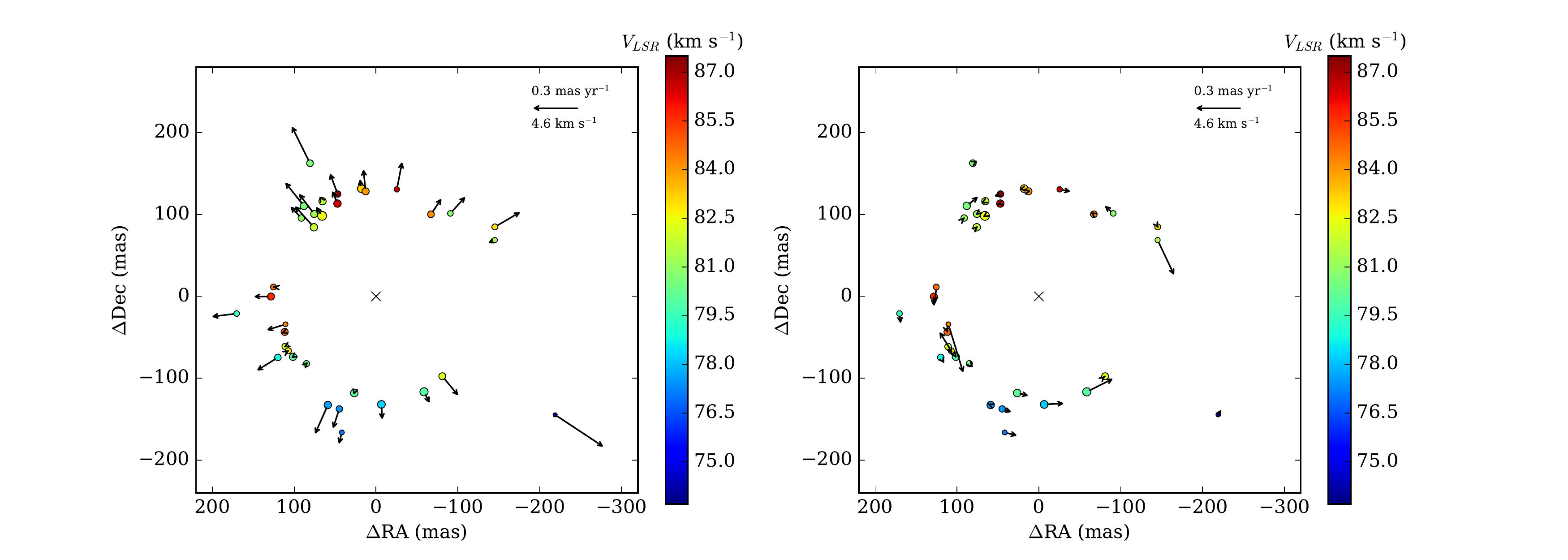}
\caption{Radial ({\it left}) and tangential ({\it right}) components of proper motion of 6.7~GHz methanol cloudlets in G23.657$-$00.127 calculated relatively to the centre of motion marked by the cross sign (top) and the centre (marked by the plus sign) of the best fitted ellipse to the cloudlet distribution at E1.
} 
\label{rotexp}  
\end{figure*}

\begin{figure}
\centering
\includegraphics[scale=0.6, trim={0 1.8cm 0 0}]{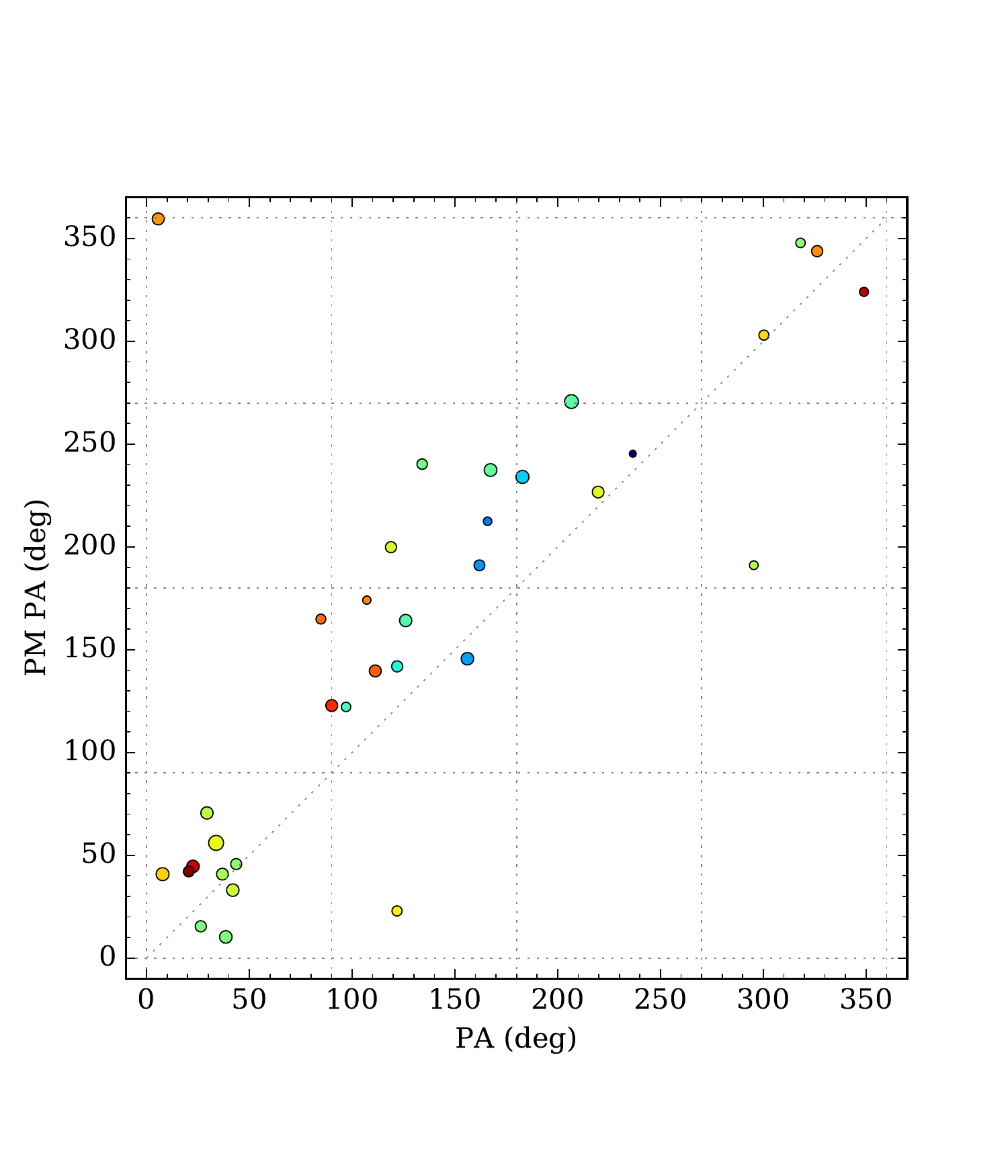}
\caption{
The position angle (north to east) of the maser cloudlet (PA) vs. the position angle (north to east) of its proper motion (PA~PM) as presented in Fig.~\ref{g23657cm} (left). The dotted lines at 0$^{\rm o}$, 90$^{\rm o}$, 180$^{\rm o}$, 270$^{\rm o}$, and 360$^{\rm o}$ are marked for clarity, as is the PA~PM = PA line.} 
\label{pmpa}  
\end{figure}

\begin{figure*}
\centering
\includegraphics[scale=0.53, trim={2.5cm 0 4.0cm 0},clip]{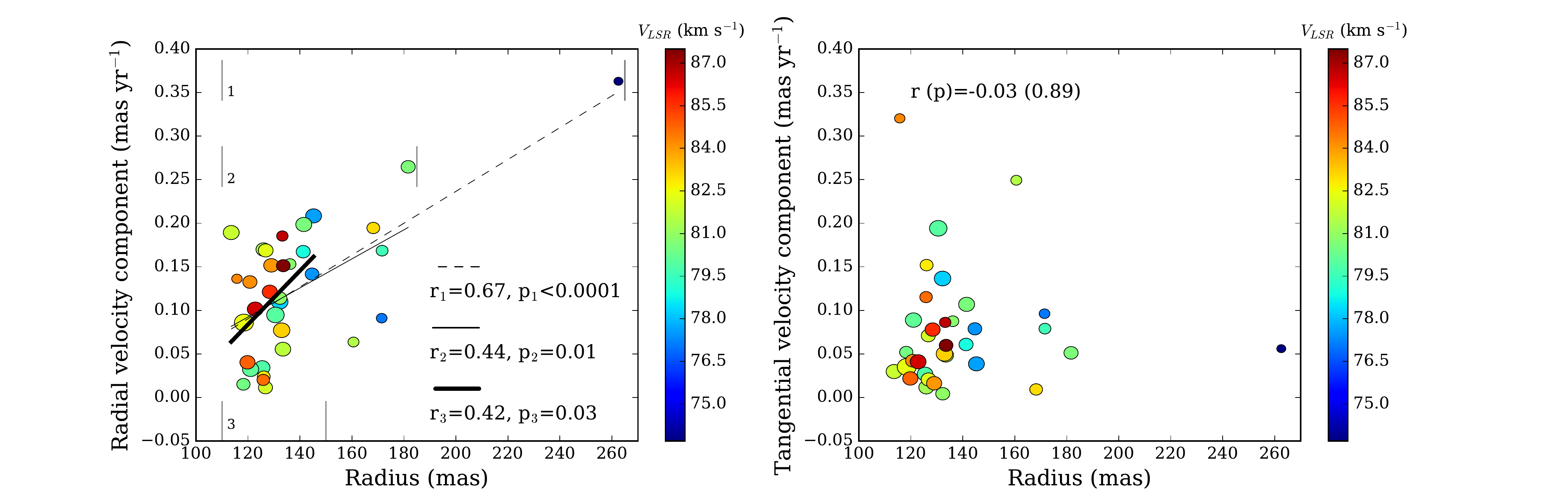}
\caption{Radial ({\it left}) and tangential ({\it right}) components of the cloudlet proper motions (Fig.~\ref{g23657cm}). The least square fits to the points and the correlation coefficients and their significance levels are displayed. Three fits are shown for the radial velocity components to verify significance of correlation when the outermost cloudlets are excluded (for clarity, each fit is presented only for the given range of considered data and vertical lines mark the radius ranges): (1) all the data points, (2) the most blue-shifted cloudlet is excluded, (3) cloudlets at radius below 150~mas only.} 
\label{rad-exp}  
\end{figure*}

 A dominating signature of radially outward moving maser cloudlets is clearly detected. We calculated their radial and tangential components relative to the centre of the ring; the results are shown in Fig.~\ref{rotexp}.
The majority of maser cloudlets show a preferentially outward, radial motion ($\sim$0.5 to
5.4~km~s$^{-1}$) in the sky. Tangential components of average value of 0.7~km~s$^{-1}$ hint at an anticlockwise rotation of cloudlets in the southern part
of the ring but they do not show any consistent rotation pattern in the northern side. A similar effect is seen when we compare the position angle (PA) of a
maser cloudlet and the position angle of its proper motion, PA~PM (Fig.~\ref{pmpa}). The blue-shifted masers, at a range of position angles from 
90$^{\rm o}$ to 270$^{\rm o}$ (north to east), show higher values of PA~PM than the PA values, indicating a non-radial component. For the red-shifted
cloudlets, the orientation of proper motions and the position angle of the cloudlets are similar (ranges from 0$^{\rm o}$ to 90$^{\rm o}$ and 270$^{\rm o}$ to 360$^{\rm o}$). There is a weak indication that the radial component of velocity increases with the cloudlet's distance from the ring centre (Fig.~\ref{rad-exp}). However,
by removing the blue-shifted cloudlet at 73.7~km~s$^{-1}$ strongly reduces the statistical significance of the correlation. The tangential velocity component
does not show any relationship with distance from the barycentre  (Fig.~\ref{rad-exp}).

\begin{figure*}
\centering
\includegraphics[width=\textwidth, trim={0cm 0 0 0},clip]{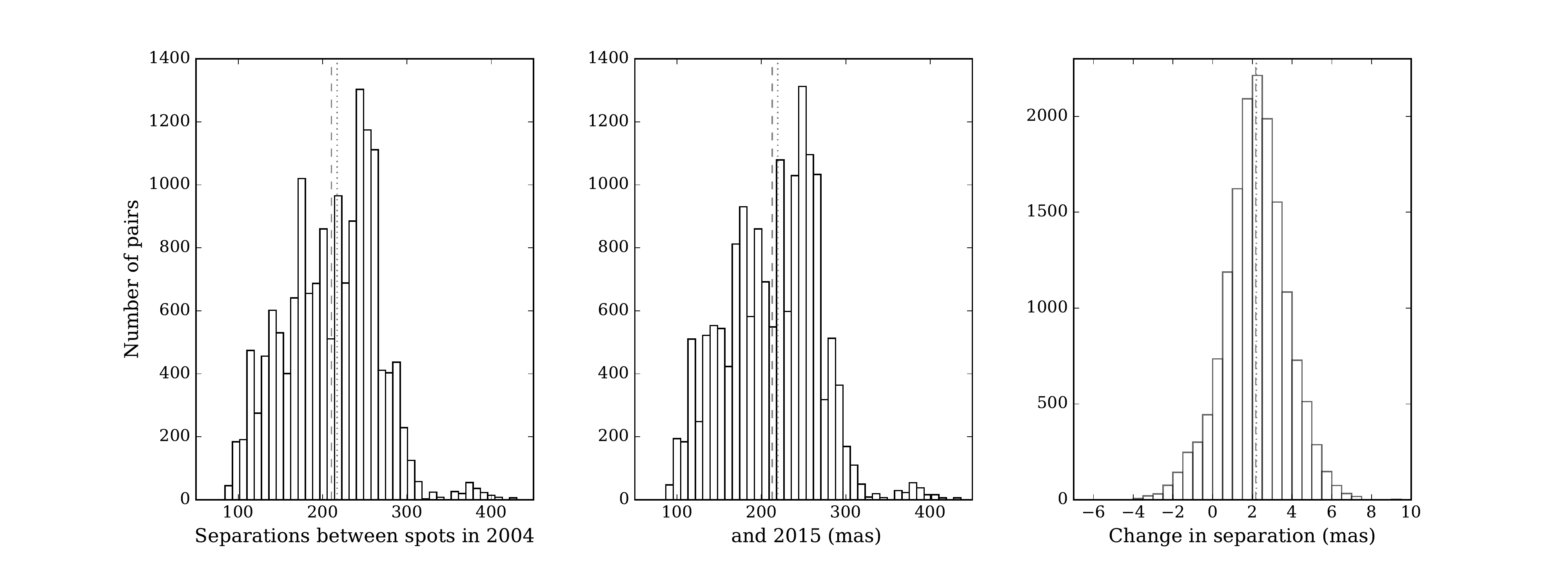}
\caption{Histograms of separations between all northern and all southern maser spot pairs in 2004 (E1) and 2015 (E3) as well as the separation increase between two epochs, respectively. The mean and median values are marked by dashed and dotted lines, respectively.} 
\label{histograms}  
\end{figure*}

The emission in the ring is clearly split between a southern and a northern part; the position angle of the cutoff line is about 80\degr~counting from north to east (Fig.~\ref{three}). In order to verify the internal motions in the ring in a way that is independent from the assumed centre, we calculated the change of separations between maser spots (not cloudlets) from these two parts between epochs E1 and E3. It also provided more statistics; the southern part consists of 136 spots, and the northern one 114 spots. For the majority of spot pairs, the separation increased (Fig.~\ref{histograms}), supporting the expansion scenario. The mean and median separations are 210~mas and 218.4~mas, and 212.7~mas and 219.3~mas at E1 and E3, respectively. The mean and median values of change in the separation after 10.4\,yr are 2.17 and 2.20~mas, respectively. Here, a 2~mas displacement over 10.3\,yr corresponds to a velocity of 2.9~km~s$^{-1}$.

The spectrum of the emission at each epoch is very complex, with several velocity
components blended together. Gaussian component fits to the spectra are shown in Fig.\,\ref{spectra}. From the first epoch (hereafter E1), 40 cloudlets are fitted with Gaussian profiles. Most of them (27) represent a single Gaussian component, while the rest are
blends of two to five components. In the second epoch (hereafter E2), there are 29 cloudlets with a single Gaussian profile and seven, three, and one cloudlets
composed of two, three, and five Gaussians, respectively. In the third epoch (hereafter E3), 35 cloudlets showed a single Gaussian velocity profile, eight double,
two triple, and one with five Gaussian components. Typical full width at half maximum (FWHM) of the best fitted Gaussian was from 0.2 to 0.4~km~s$^{-1}$.
Gaussian fits were not found for seven, eleven, and twelve groups of spots at E1, E2, and E3, respectively. The parameters of the best Gaussian fits for the 34
persistent cloudlets at each epoch are listed in Table~\ref{clouds}.

\section{Discussion}
We discuss two major aspects of the G23.657$-$00.127 methanol maser ring; the variability of discrete cloudlets and the origin of its structure.
\subsection{Variability}
\begin{figure*}
\centering
\includegraphics[scale=0.45, trim={2cm 2cm 0 2cm},clip]{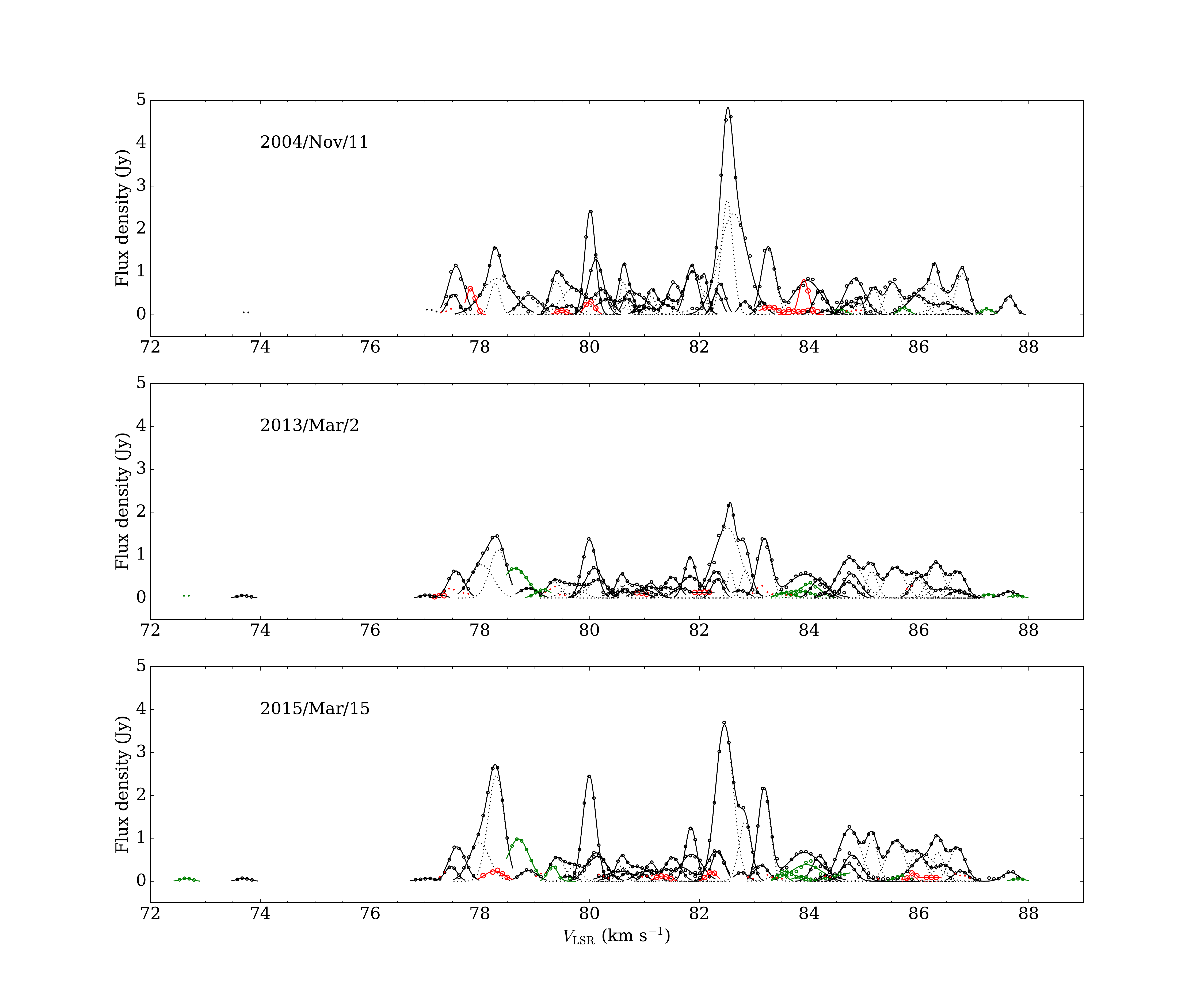}
\caption{Three epoch EVN data. 
Gaussian component fits to the EVN spectrum at each epoch. The dots indicate the emission of individual spots. The lines represent the Gaussian fits of individual cloudlets. The cloudlets persisted in all the three epochs are marked in black (Table~\ref{clouds}). Those that appeared at two out of three epochs are marked in green, and those appearing only at one epoch in red.}
\label{spectra}
\end{figure*}

The 6.7\,GHz methanol maser spectrum of the target remains constant within 10\% over nearly 20\,yr (Szymczak et al.~\cite{sz18}). However, our present EVN
observations revealed significant changes in the intensity of individual cloudlets over the whole range of velocity (Fig.~\ref{spectra}), which were not detected in the poorer signal-to-noise ratio observations and without any possibility of distinction of blended spectral features. Nine cloudlets consisting of at least two
Gaussian components showed complex structures at all three epochs. The most complex was cloudlet 31 (Table~\ref{clouds}), which had five Gaussian profiles, but their
intensity varied by 20-50\% on timescales of 2$-$10~yr. There were 10 to 14 cloudlets of brightness below 0.4~Jy~beam$^{-1}$  that appeared at one epoch only and
nine cloudlets were seen at any two out of three epochs. The appearance and disappearance of cloudlets does not show any clear dependence on the location in the ring. We estimated the average lifetime of an individual maser cloudlet in the source from the percentage of cloudlets that either appear or disappear over the intervals of 8.3 and 10.3\,yr. At E2 and E3, the mean numbers of new and vanishing cloudlets were 12 and 14, respectively, relative to that seen at E1. This implies that about 20 and 24\% of cloudlets appeared or disappeared at the respective epochs for an average lifetime of $\sim$40\,yr. This crude estimate is a factor of four lower than that obtained from a statistical analysis of maser features from single dish spectra for 21 targets observed at two epochs spanned by a decade (Ellingsen~\cite{e07}). This shorter lifetime of maser cloudlets can be related to the specific case of our target, for example related to internal instabilities of the masing region, which may reduce the velocity coherence along the path of maser rays. However, it is predominantly affected by different statistics and by differences in the interferometric and single-dish data. Higher resolution observations with a possibility of distinction of blended, weak spectral features are particularly valuable in such studies.

The maser emission is heavily resolved out with the EVN telescope;
more than 70\% of the flux is missing when compared to that from the single-dish observations.
This indicates that the majority of the emission may come from extended regions of low brightness. The width of the ring as measured with the EVN was 95\,au (Bartkiewicz et al.~\cite{b05}) and has been stable in E2 and E3, thus, at a distance of 405\,au from the central star, we estimate that the tangential mean path length of the maser rays is $\sim$400\,au. This is still much less than a typical model assumption of 6700\,au (Cragg et al.~\cite{c02}). It is also at the lower end of the maser path lengths derived
from the statistical analysis of the size of 60 sources observed using the EVN,
with a mean value of $\sim$1000\,au (Sarniak et al.~\cite{sa18}). Such a relatively short path of maser amplification can be one of the reasons for significant variability of several cloudlets in the source.

\subsection{Where does the expanding maser ring come from?}

\begin{figure}
\centering
\includegraphics[scale=0.35, trim={1cm 0cm 0cm 0cm},clip]{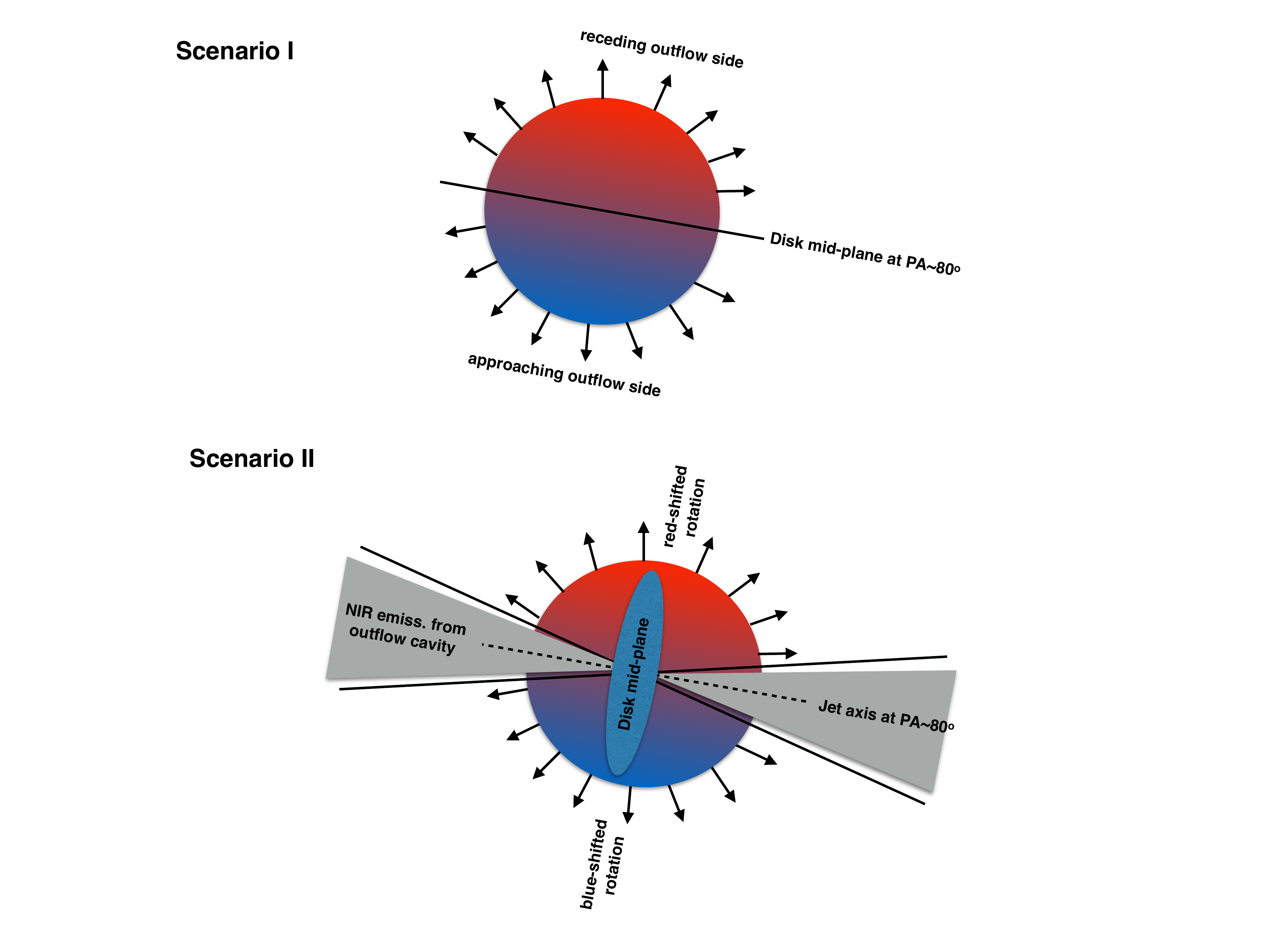}
\caption{Schematic models of maser expansion related to a sphere-like outflow (Scenario~I) or tracing a wide angle wind at the base of the protostellar jet (Scenario~II).} 
\label{scheme}  
\end{figure}

The present study provides the first constraints on the origin of the methanol maser ring in G23.657$-$00.127. The maser emission arises from an
almost circular structure of radius 405\,au with two gaps along the direction at position angle $\sim$80\degr; the velocity field traced by the masers reveals an outward radial motion with a mean velocity of 3.2~km~s$^{-1}$. This spatio-kinematical structure might resemble that predicted by MHD modellings in the inner circumstellar 
envelope of very young stars (Seifried et al.~\cite{se12}, Kuiper et al.~\cite{k15}, Matsushita et al.~\cite{ma18}). Simulations of outflows formed during the collapse
of massive (100$M_{\odot}$) cloud cores reveal that the morphology of outflows launched from the disks by magneto-centrifugal acceleration strongly depends on
the magnetic field strength. For instance, for slightly supercritical values of the mass-to-flux ratios $\mu\approx 5$, that is to say, for strong magnetic fields (B$\approx$0.7\,mG), the outflow has a sphere-like morphology at an early stage of evolution ($\sim$4000~yr), expanding with roughly the same velocity up to 5~km~s$^{-1}$ in all directions (Seifried et al.~\cite{se12}, their Fig.~7).

In the following, we highlight pros and cons of this scenario (Scenario~I, Fig.~\ref{scheme}) applied to interpreting the maser ring. The source G23.657$-$00.127, similarly to other methanol maser
sources showing a ring-like morphology, is not associated with bright H~{\small II} emission at cm wavelengths (Bartkiewicz et al.~\cite{b09}), nor with water masers
(Bartkiewicz et al.~\cite{b11}, Urquhart et al.~\cite{u11}) or hydroxyl masers (Szymczak \& G\'erard \cite{sz04}). This evidence would suggest that G23.657$-$00.127 is at an early stage of evolution,
in agreement with the simulations by Seifried et al.~(\cite{se12}). On the other hand, De Buizer et al. ~(\cite{d12}) detected near- (NIR) and mid-infrared (MIR) extended
emission towards the ring; if this emission were directly associated with the source driving the methanol masers, then the assumption of a star at an early stage of evolution
would be questionable. Furthermore, in Fig.~7 of Seifried et al.~(\cite{se12}) the gas density along the outer front of the expansion changes significantly, and a spherical distribution of maser cloudlets might be difficult to see. 
Nevertheless, both the values of mass outflow rate ($10^{-4}$~M$_{\odot}$~yr$^{-1}$) and momentum rate ($10^{-4}$~M$_{\odot}$~km~s$^{-1}$yr$^{-1}$) calculated by Seifried et al.~(\cite{se12}) are consistent with those expected for an early stage of star formation like in G023.01$-$00.41 (Sanna et al.~\cite{s14}) or AFLG~5142 (Goddi et al.~\cite{g11}). Similar calculations for masers in G23.657$-$00.127 as presented in details in Goddi et al.~(\cite{g11}), who  considered a sphere-like outflow, lead to values of the mass-loss rate of $3.9\times10^{-3} n_8 M_\odot$~yr$^{-1}$ and the momentum rate of $6.3\times10^{-2} n_8 M_\odot$~yr$^{-1}$~km~s$^{-1}$ of the molecular flow, where $n_8$ is the ambient volume density in units of $10^8$~cm$^{-3}$. This calculation assumes the average distance of masers from the protostar of 405~AU and the average maser velocity of 16~km~s$^{-1}$. 

Following Scenario~I, we note that a peculiar feature of the maser distribution is the emission-free part of the maser ring along a line at PA\,$\sim$80\degr
(N to E) (Fig.~\ref{three}). This region might mark the mid-plane of a circumstellar disc, and the masers would expand above and below this plane. In this respect, if other methanol maser rings
were also related to a very young (almost spherical) outflow, they would also expand with constant eccentricity, showing a gap between red- and blue-shifted cloudlets. This
prediction can be verified observationally in the near future. Nevertheless, the orientation of this emission-free region is also consistent with the elongation of the NIR emission
reported by De Buizer et al. ~(\cite{d12}, their Fig.~4), which was interpreted as a tracer of an underlying jet, inclined with respect to the line of sight. This evidence opens
up the possibility of Scenario~II (Fig.~\ref{scheme}): the HMYSO would still lie at the centre of the methanol maser ring, but the masers would now be expanding each side of the jet axis, and
would trace a wide angle wind at the base of the protostellar jet. In this scenario, the proper  motion and line-of-sight velocity of the masers would be tracing a combination of rotation around and expansion along the jet axis, similarly to what was observed in the maser source G23.01$-$0.41 (Sanna et al.~\cite{s10b}).

In order to verify whether one of the above scenarios applies to G23.657$-$00.127, or if another interpretation is required, we need complementary observations of
thermal tracers at high angular resolution ($\sim$0\farcs1). In this respect, we have started a sensitive ($\sim \mu$Jy) search of radio continuum emission with the
Very Large Array. These observations will be able to detect the ionized gas emission excited by any B-type young star associated with the masers, such that as due to radio thermal
jets or hyper-compact H~{\small II} regions, and will allow us to pinpoint the driving source of the masers at a resolution comparable with the extent of the maser ring.

\section{Conclusions}
A detailed study of the proper motions of the methanol maser ring G23.657$-$00.127 clearly indicates that the underlying structure is expanding at a mean velocity
of 3.2~km~s$^{-1}$. The radial components of proper motions dominate over the tangential components. While the overall morphology of the ring has been stable in time,
individual maser cloudlets have varied in brightness, suggesting internal instabilities of the masing region. At present, we can speculate about two possible scenarios
that might explain the spatio-kinematical structure revealed by the masers. Both interpretations require the presence of a wide-angle wind from a young star, but in one case, the wind would be found at the base of a protostellar jet detected in the near-infrared. Complementary observations of thermal tracers at high angular resolution are needed to discern between the different scenarios.

\begin{acknowledgements}
AB, MS, PW acknowledge support from the National Science Centre, Poland through grant 2016/21/B/ST9/01455. We thank Prof.~Krzysztof Goździewski for fruitful discussions about the error estimation and the bootstrap method. We also thank the anonymous referee for detailed check of the manuscript. The research leading to these results has received funding from the European Commission Seventh Framework Programme (FP/2007-2013) under grant agreement No. 283393 (RadioNet3).
\end{acknowledgements}


\Online
\appendix
\section{Table and Figures}

\onecolumn
\begin{longtable}{c r r c c c c c c c c}
\caption{\label{clouds} Parameters of 6.7~GHz methanol maser cloudlets that appeared at all three epochs and were used for proper motion studies. Coordinates ($\Delta$RA,$\Delta$Dec) are relative to the motion centre as in Fig.~\ref{g23657cm}. If possible, Gaussian velocity profiles are characterised by the velocity (V$_{\rm fit}$), full
width at half maximum (FWHM), and flux density (S$_{\rm fit}$) of the maser cloudlet. The relative proper motion of cloudlets estimated by linear fitting are also listed ($\mu_{\rm RA}$, $\mu_{\rm Dec}$).}\\
\hline\hline
 & $\Delta$RA & $\Delta$Dec & V$_{\rm p}$ & V$_{\rm fit}$ & FWHM & S$_{\rm p}$ & S$_{\rm fit}$ & L$_{\rm proj}$ & $\mu_{\rm RA}$ & $\mu_{\rm Dec}$\\
& (mas) & (mas) & (km~s$^{-1}$) & (km~s$^{-1}$) & (km~s$^{-1}$) & (Jy~beam$^{-1}$)& (Jy~beam$^{-1}$) & (AU) & (mas~yr$^{-1}$) & (mas~yr$^{-1}$)\\
 & & & & & & & & & (km~s$^{-1}$) & (km~s$^{-1}$)\\
\hline                    
\endfirsthead
\caption{continued.}\\
\hline\hline
 & $\Delta$RA & $\Delta$Dec & V$_{\rm p}$ & V$_{\rm fit}$ & FWHM & S$_{\rm p}$ & S$_{\rm fit}$ & L$_{\rm proj}$ & $\mu_{\rm RA}$ & $\mu_{\rm Dec}$ \\
& (mas) & (mas) & (km~s$^{-1}$) & (km~s$^{-1}$) & (km~s$^{-1}$) & (Jy~beam$^{-1}$)& (Jy~beam$^{-1}$) & (AU) & (mas~yr$^{-1}$) & (mas~yr$^{-1}$)\\
 & & & & & & & & & (km~s$^{-1}$) & (km~s$^{-1}$)\\
\hline
\endhead
\hline
\endfoot
\hline
\endlastfoot
\multicolumn{10}{l}{\bf Cloudlet 1}\\
E1 & -253.533 & -152.352 & 73.70 & & & 0.057 & & 1.6 & -0.33$\pm$0.05  & -0.16$\pm$0.03 \\ 
E2 & -256.310 & -155.497 & 73.66 & 73.69 & 0.27 & 0.061 & 0.059 & 6.2 &(-5.0$\pm$0.8) & (-2.4$\pm$0.4) \\ 
E3 & -257.195 & -154.774 & 73.67 & 73.70 & 0.27 & 0.071 & 0.069 & 4.4 \\
\multicolumn{10}{l}{\bf Cloudlet 2}\\
E1 & 6.803 & -173.999 &  77.03 & & & 0.123 & & 6.3  & -0.07$\pm$0.03  & -0.11$\pm$0.01 \\ 
E2 & 5.433 & -176.891 & 77.00 & 77.04 & 0.29 & 0.066 & 0.070 & 8.4 & (-1.1$\pm$0.5) & (-1.7$\pm$0.2)\\ 
E3 & 6.598 & -176.101 & 77.09 & 77.04 & 0.44 & 0.066 & 0.060 & 18.0 \\
\multicolumn{10}{l}{\bf Cloudlet 3}\\
E1 & 26.630 & -141.485 & 77.56 & 77.57 & 0.34 & 1.152 & 1.129 & 20.5 & 0.12$\pm$0.02  & -0.18$\pm$0.05 \\
E2 & 26.888 & -145.149 & 77.53 & 77.57 & 0.34 & 0.622 & 0.631 & 18.7& (1.8$\pm$0.3) & (-2.7$\pm$0.7) \\ 
E3 & 26.845 & -144.007 & 77.53 & 77.59 & 0.36 & 0.791 & 0.802 & 25.6\\
\multicolumn{10}{l}{\bf Cloudlet 4}\\
E1 & 9.339 & -145.446 & 77.56 & 77.51 & 0.25 & 0.452 & 0.481 & 9.6 & -0.03$\pm$0.01  & -0.16$\pm$0.01 \\ 
E2 & 9.973 & -148.564 & 77.44 & & & 0.220 & & 11.6 & (-0.5$\pm$0.1) & (-2.4$\pm$0.1)\\ 
E3 & 10.024 & -147.938 & 77.44 & 77.47 & 0.26 & 0.326 & 0.344 & 14.8 \\ 
\multicolumn{10}{l}{\bf Cloudlet 5}\\
E1 & -39.464 & -140.038 & 78.27 & 78.28 & 0.21 & 1.556 & 0.741 & 67.2 & -0.14$\pm$0.01  & -0.10$\pm$0.01 \\ 
E1 & &  &  & 78.32 & 0.68 & & 0.848 & & (-2.1$\pm$0.1 ) & (-1.5$\pm$0.1)\\
E2 & -40.751 & -142.033 & 78.32 & 78.03 & 0.50 & 1.445 & 0.772 & 16.2 \\
E2 & & & & 78.33 & 0.37 & & 1.118 &  \\ 
E3 & -40.972 & -141.999 & 78.32 & 77.99 & 0.44 & 2.643 & 0.897 & 18.0 \\ 
E3 & & & & 78.30 & 0.35 & & 2.469 & \\
\multicolumn{10}{l}{\bf Cloudlet 6}\\
E1 & 86.212 & -83.371 & 78.88 & 78.90 & 0.44 & 0.515 & 0.460 & 23.2 & 0.11$\pm$0.01  & -0.14$\pm$0.01 \\ 
E2 & 86.009 & -85.654 & 78.93 & 78.88 & 0.40 & 0.226 & 0.229 & 14.6& (1.6$\pm$0.1) & (-2.1$\pm$0.1) \\ 
E3 & 87.093 & -86.333 & 78.85 & 78.89 & 0.36 & 0.263 & 0.255 & 18.9 \\ 
\multicolumn{10}{l}{\bf Cloudlet 7}\\
E1 & 67.451 & -84.341 & 79.41 & 79.39 & 0.27 & 0.996 & 0.780 & 19.6 & 0.01$\pm$0.01  & -0.04$\pm$0.01 \\ 
E1 & & & & 79.70 & 0.51 & & 0.612 &  & (0.2$\pm$0.2) & (-0.6$\pm$0.2)\\
E1 & & & & 80.25 & 0.36 & & 0.568 &  \\
E2 & 65.743 & -83.326 & 80.16 & 79.35 & 0.31 & 0.429 & 0.377 & 19.2 \\ 
E2 &  &  &  & 79.69 & 0.41 & & 0.308 &  \\
E2 &  &  &  & 80.16 & 0.41 & & 0.415 &  \\
E3 & 65.861 & -82.886 & 80.17 & 79.40 & 0.33 & 0.589 & 0.541 & 18.8 \\
E3 & & &  & 79.71 & 0.29 & & 0.332 &  \\
E3 & & &  & 80.15 & 0.43 & & 0.581 &  \\
\multicolumn{10}{l}{\bf Cloudlet 8\footnote{Weaker Gaussian feature was used for proper motions.}}\\
E1 & 152.045 & -20.684 & 79.32 & 79.31 & 0.21 & 0.224 & 0.215 & 87.2 &  \\ 
E1 & 134.537 & -29.093 & 79.67 & 79.64 & 0.32 &0.205  & 0.204 & &  0.15$\pm$0.05  & -0.10$\pm$0.01\\
E2 & 135.040 & -31.685 & 79.63 &  & & 0.103 & & 10.3 & (2.3$\pm$0.7) & (-1.5$\pm$0.2) \\ 
E3 & 136.778 & -31.472 & 79.64 & 79.60 & 0.38 & 0.108 & 0.111 & 13.3 \\ 
\multicolumn{10}{l}{\bf Cloudlet 9}\\
E1  & -93.127 & -124.285 & 80.02 & 80.01 & 0.23 & 2.406 & 2.437 & 8.7 & -0.22$\pm$0.01  & 0.01$\pm$0.05 \\ 
E2  & -95.110 & -125.658 & 79.98 & 80.00 & 0.30 & 1.379 & 1.355 & 4.9 & (-3.3$\pm$0.1) & (0.1$\pm$0.8)\\ 
E3  & -95.544 & -125.577 & 79.99 & 80.00 & 0.28 & 2.443 & 2.461 & 4.2 \\ 
\multicolumn{10}{l}{\bf Cloudlet 10}\\
E1  & -7.721 & -125.195 & 80.11 & 80.12 & 0.31 & 1.316 & 1.284 & 20.4 & -0.08$\pm$0.01  & -0.05$\pm$0.01 \\ 
E2  & -8.968 & -128.111 & 80.07 & 80.09 & 0.38 & 0.708 & 0.695 & 18.7 & (-1.2$\pm$0.1) & (-0.8$\pm$0.1)\\ 
E3  & -9.382 & -127.429 & 80.08 & 80.12 & 0.42 & 0.668 & 0.661 & 20.9 \\ 
\multicolumn{10}{l}{\bf Cloudlet 11}\\
E1 & 50.535 & -90.492 & 80.38 & 80.34 & 0.55 & 0.355 & 0.360 & 10.2 & -0.04$\pm$0.02  & -0.03$\pm$0.02 \\ 
E1 & & & & 80.71 & 0.24 & & 0.256 & & (-0.7$\pm$0.3) & (-0.4$\pm$0.2)\\
E2 & 49.989 & -91.483 & 80.60 & 80.56 & 0.55 & 0.145 & 0.146 & 4.2 \\ 
E3 & 49.811 & -90.444 & 80.61 & 80.56 & 0.65 & 0.263 & 0.232 & 8.9 \\ 
\multicolumn{10}{l}{\bf Cloudlet 12}\\
E1 & 45.898 & 154.204 & 80.73 & 80.71 & 0.24 & 0.537 & 0.541 & 12.2 & 0.08$\pm$0.03  & 0.26$\pm$0.01 \\ 
E2 & 47.109 & 155.542 & 80.60 & 80.64 & 0.23 & 0.202 & 0.217 & 4.8 & (1.1$\pm$0.5) & (3.9$\pm$0.1)\\ 
E3 & 46.225 & 155.697 & 80.69 & 80.68 & 0.25 & 0.192 & 0.195 & 6.4 \\ 
\multicolumn{10}{l}{\bf Cloudlets 13 and 14\footnote{Both Gaussian features were used for proper motions.}}\\
E1 & 53.535 & 101.834 & 80.64 & 80.62 & 0.16 & 1.178 & 0.768 & 77.5 & 0.04$\pm$0.01$^a$  & 0.22$\pm$0.03$^a$\\ 
E1 & 55.143 & 93.753 & 80.81 & 80.78 & 0.63 & & 0.513 & & 0.08$\pm$0.02$^b$  & 0.08$\pm$0.01$^b$ \\
E2 & 53.763 & 101.775 & 80.60 & 80.58 & 0.16 & 0.562 & 0.300 & 78.8 & (0.6$\pm$0.1)$^a$ & (3.4$\pm$0.5)$^a$\\ 
E2 & 57.322 & 88.883 & 80.86  & 80.75 & 0.70 & & 0.318 & & (1.2$\pm$0.3)$^b$ & (1.2$\pm$0.1)$^b$\\
E3 & 53.919 & 102.736 & 80.61 & 80.58 & 0.19 & 0.600 & 0.357 & 101.2 \\
E3 & 57.125 & 90.007 & 80.87 & 80.82 & 0.67 & & 0.345 & \\
\multicolumn{10}{l}{\bf Cloudlet 15}\\
E1  & -125.576 & 93.581 & 80.99 & 80.96 & 0.28 & 0.214 & 0.223 & 3.5 & -0.03$\pm$0.02  & 0.17$\pm$0.05 \\ 
E2  & -125.793 & 92.900 & 80.86 & 80.90 & 0.32 & 0.187 & 0.190 & 5.3 & (-0.5$\pm$0.3) & (2.5$\pm$0.7)\\ 
E3  & -126.072 & 94.390 & 80.96 & 80.92 & 0.29 & 0.193 & 0.204 & 2.9 \\ 
\multicolumn{10}{l}{\bf Cloudlet 16}\\
E1 & -179.205 & 59.177 & 81.61 & & & 0.249 & & 6.1 & -0.05$\pm$0.04  & -0.24$\pm$0.09 \\ 
E2 & -180.428 & 57.766 & 81.48 & & & 0.100& & & (-0.8$\pm$0.6) & (-3.7$\pm$1.4)\\ 
E3 & -180.471 & 57.523 & 81.57 & & & 0.093 & &  1.8 \\ 
\multicolumn{10}{l}{\bf Cloudlet 17}\\
E1 & 38.705 & 91.985 & 81.52 & 81.04 & 0.37 & 0.764 & 0.161 & 33.9 & 0.11$\pm$0.01  & 0.13$\pm$0.01  \\ 
E1 & & & & 81.54 & 0.31 & & 0.738 & & (1.7$\pm$0.2) & (2.0$\pm$0.1)\\
E2 & 39.022 & 91.172 & 81.48 & 81.07 & 0.40 & 0.481 & 0.269 & 31.7 \\ 
E2 & & & & 81.51 & 0.30 & & 0.480 & \\
E3 & 39.398 & 91.892 & 81.48 & 81.08 & 0.32 & 0.546 & 0.261 & 30.1 \\ 
E3  & & & & 81.51 & 0.33 & & 0.555 & \\
\multicolumn{10}{l}{\bf Cloudlet 18}\\
E1 & 41.846 & 77.244 & 81.87 & 81.86 & 0.27 & 1.153 & 1.149 & 32.4 & 0.11$\pm$0.01  & 0.16$\pm$0.01 \\ 
E2 & 42.343 & 76.458 & 81.83 & 81.83 & 0.26 & 0.938 & 0.950 & 8.6 & (1.6$\pm$0.2 ) & (2.4$\pm$0.1)\\ 
E3 & 42.456 & 77.550 & 81.83 & 81.85 & 0.24 & 1.228 & 1.265 & 8.4 \\ 
\multicolumn{10}{l}{\bf Cloudlet 19}\\
E1 & 29.657 & 107.075 & 81.87 & 81.42 & 0.27 & 1.006 & 0.378 & 41.4 & 0.07$\pm$0.05  & 0.02$\pm$0.04 \\ 
E1 & & & & 81.88 & 0.37 &  & 1.011 & & (1.1$\pm$0.8) & (0.3$\pm$0.6)\\
E1 & & & & 82.10 & 0.12 &  & 0.561 &  \\
E2 & 30.247 & 106.219 & 81.83 & 81.37 & 0.24 & 0.505 & 0.226 & 31.4 \\ 
E2 & & & & 81.84 & 0.46 & & 0.496 & \\
E3 & 30.254 & 107.307 & 81.83 & 81.37 & 0.25 & 0.598 & 0.252 & 18.9 \\
E3 & & & & 81.86 & 0.46 & & 0.629 & \\
\multicolumn{10}{l}{\bf Cloudlet 20}\\
E1 & 72.954 & -74.195 & 82.84 & 82.82 & 0.23 & 0.304 & 0.307 & 2.2 & 0.06$\pm$0.01  & 0.15$\pm$0.06 \\ 
E2 & 73.199 & -73.980 & 82.71 & 82.73 & 0.36 & 0.179 & 0.181 & 3.7& (0.9$\pm$0.1 ) & (2.2$\pm$0.9) \\ 
E3 & 73.004 & -73.894 & 82.80 & 82.76 & 0.28 & 0.194 & 0.206 & 3.5 \\ 
\multicolumn{10}{l}{\bf Cloudlet 21}\\
E1 & 76.096 & -68.423 & 82.31 & 82.32 & 0.21 & 0.511 & 0.515 & 4.8 & -0.02$\pm$0.02  & -0.07$\pm$0.01  \\ 
E2 & 75.413 & -71.325 & 82.27 & 81.68 & 0.43 & 0.606 & 0.224 & 13.4 & (-0.3$\pm$0.3) & (-1.0$\pm$0.1)\\ 
E2 & & & & 82.31 & 0.34 & & 0.619 & \\
E3 & 75.269 & -70.270 & 82.27 & 81.68 & 0.42 & 0.703 & 0.320 & 14.7 \\ 
E3 & & & & 82.30 & 0.36 & & 0.700 & \\
\multicolumn{10}{l}{\bf Cloudlet 22}\\
E1  & -115.358 & -105.101 & 82.40 & 82.37 & 0.24 & 0.709 & 0.727 & 4.1 & -0.12$\pm$0.01  & -0.12$\pm$0.01 \\ 
E2 & -116.431 & -108.533 & 82.36 & 82.33 & 0.23 & 0.432 & 0.449 & 6.0 & (-1.9$\pm$0.1) & (-1.8$\pm$0.1)\\ 
E3 & -116.699 & -107.896 & 82.36 & 82.34 & 0.27 & 0.683 & 0.696 & 29.1 \\
\multicolumn{10}{l}{\bf Cloudlet 23}\\
E1 & 31.639 & 88.650 & 82.57 & 82.51 & 0.25 & 4.619 & 2.665 & 38.3 & 0.07$\pm$0.02  & 0.05$\pm$0.01\\ 
E1 & & & & 82.62 & 0.61 & & 2.359 &  & (1.1$\pm$0.3) & (0.8$\pm$0.2) \\
E2 & 31.752 & 86.848 & 82.53 & 82.50 & 0.57 & 2.152 & 1.637 & 37.5 & \\ 
E2 &  &  &  & 82.57 & 0.14 &  & 0.642 &  \\
E2 &  &  &  & 82.84 & 0.21 &  & 0.632 &  \\
E3 & 32.219 & 89.523 & 82.45 & 82.46 & 0.37 & 3.699 & 3.641 & 34.0 \\ 
E3 & & & & 82.83 & 0.27 & & 1.381 & \\
\multicolumn{10}{l}{\bf Cloudlet 24}\\
E1  & -179.404 & 76.999 & 83.10 & 83.14 & 0.22 & 0.293 & 0.310 & 4.0 & -0.17$\pm$0.06  & 0.11$\pm$0.04 \\ 
E2  & -181.984 & 76.686 & 83.15 &  & & 0.288 & & 2.6 & (-2.6$\pm$0.9) & (1.7$\pm$0.7)\\ 
E3  & -181.533 & 77.319 & 83.15 & 83.13 & 0.31 & 0.371 & 0.374 & 4.7 \\ 
\multicolumn{10}{l}{\bf Cloudlets 25 and 26\footnote{Both Gaussian features were used for proper motions.}}\\
E1 & -16.208 & 123.792 & 83.28 & 83.25 & 0.30 & 1.519 & 1.577 & 35.9 & 0.06$\pm$0.02$^a$  & 0.07$\pm$0.01$^a$ \\ 
E1 & -21.252 & 120.618 & 83.98 & 83.97 & 0.59 & 0.844 & 0.799 & & 0.00$\pm$0.01$^b$  & 0.15$\pm$0.02$^b$ \\
E2  & -15.348 & 122.310 & 83.15 & 83.18 & 0.29 & 1.376 & 1.363 & 37.6 & (0.9$\pm$0.2)$^a$ & (1.1$\pm$0.1)$^a$\\ 
E2 & -21.334 & 119.999 & 83.94 & 83.93 & 0.73 & 0.591 & 0.554 & & (0.0$\pm$0.1)$^b$ & (2.3$\pm$0.3)\\
E3  & -15.293 & 122.870 & 83.15 & 83.18 & 0.26 & 2.160 & 2.151 & 34.8 \\
E3 & -21.352 & 121.319 & 83.94 & 83.91 & 0.72 & 0.691 & 0.674 & \\
\multicolumn{10}{l}{\bf Cloudlet 27}\\
E1 & -102.322 & 92.561 & 84.24 & 84.21 & 0.27 & 0.550 & 0.572 & 9.8 & -0.04$\pm$0.03  & 0.13$\pm$0.03 \\ 
E2 & -102.881 & 91.324 & 84.20 & 84.19 & 0.34 & 0.441 & 0.442 & 9.9 & (-0.6$\pm$0.4) & (2.0$\pm$0.5)\\ 
E3 & -102.858 & 92.632 & 84.20 & 84.19 & 0.31 & 0.592 & 0.587 & 18.2 \\ 
\multicolumn{10}{l}{\bf Cloudlet 28}\\
E1 & 75.805 & -42.352 & 84.33 & 84.31 & 0.23 & 0.107 & 0.110 & 4.3 & 0.03$\pm$0.03  & -0.35$\pm$0.02 \\ 
E2 & 76.201 & -47.031 & 84.29 & 84.27 & 0.24 & 0.127 & 0.128 & 2.1 & (0.5$\pm$0.5) & (-5.3$\pm$0.3)\\
E3 & 76.554 & -46.609 & 84.29 & 84.30 & 0.23 & 0.138 & 0.138 & 2.4 \\
\multicolumn{10}{l}{\bf Cloudlet 29}\\
E1 & 77.392 & -51.483 & 84.86 & 84.83 & 0.42 & 0.838 & 0.848 & 25.1 & 0.03$\pm$0.02  & -0.03$\pm$0.01 \\
E2 & 77.077 & -52.809 & 84.73 & 84.77 & 0.38 & 0.586 & 0.563 & 12.0 & (0.4$\pm$0.3) & (-0.5$\pm$0.1)\\ 
E3 & 77.121 & -51.754 & 84.73 & 84.78 & 0.39 & 0.656 & 0.610 & 25.4 \\
\multicolumn{10}{l}{\bf Cloudlet 30}\\
E1 & 93.386 & -3.914 & 84.95 & 84.69 & 0.28 & 0.395 & 0.242 & 43.5 &0.04$\pm$0.05  & -0.12$\pm$0.05 \\ 
E1 & & & & 84.93 & 0.12 & & 0.400 & & (0.5$\pm$0.8) & (-1.8$\pm$0.8)\\
E2 & 90.422 & 0.116 & 84.73 & 84.71 & 0.33 & 0.379 & 0.368 & 10.7 \\
E3 & 90.266 & 1.301 & 84.73 & 84.71 & 0.32 & 0.418 & 0.401 & 20.4 \\
\multicolumn{10}{l}{\bf Cloudlet 31}\\
E1 & 92.768 & -6.569 & 85.56 & 84.84 & 0.91 & 0.821 & 0.351 & 20.7 & +0.12$\pm$0.01  & -0.08$\pm$0.01 \\ 
E1 &  &  &  & 85.18 & 0.24 &  & 0.379 &  & (1.8$\pm$0.1) & (-1.2$\pm$0.1)\\
E1 &  &  &  & 85.52 & 0.30 &  & 0.733 &  \\
E1 &  &  &  & 85.94 & 0.40 &  & 0.437 &  \\
E1 &  &  &  & 86.47 & 0.58 &  & 0.253 &  \\
E2 & 97.019 & -11.151 & 84.73 & 85.14 & 0.24 & 0.967 & 0.618 & 23.6 \\
E2 &  &  & & 84.74 & 0.51 &  & 0.921 &  \\
E2 &  &  & & 85.56 & 0.41 &  & 0.710 &  \\
E2 &  &  & & 85.97 & 0.33 &  & 0.532 &  \\
E2 &  &  & & 86.51 & 0.62 &  & 0.211 &  \\
E3 & 97.693 & -10.776 & 84.73 & 84.74 & 0.47 & 1.268 & 1.220 & 26.8 \\
E3 &  &  &  & 85.15 & 0.26 &  & 0.982 &  \\
E3 &  &  &  & 85.57 & 0.39 &  & 0.932 &  \\
E3 &  &  &  & 85.97 & 0.35 &  & 0.656 &  \\
E3 &  &  &  & 86.45 & 0.41 &  & 0.381 &  \\
\multicolumn{10}{l}{\bf Cloudlet 32}\\
E1 & 13.707 & 103.628 & 86.26 & 86.29 & 0.15 & 1.174 & 0.500 & 42.4 & 0.08$\pm$0.01  & 0.08$\pm$0.01 \\ 
E1 &  &  &  & 86.23 & 0.72 &  & 0.736 &  & (1.2$\pm$0.1 ) & (1.2$\pm$0.1)\\
E1 &  &  &  & 86.79 & 0.28 &  & 0.955 &  \\
E2 & 13.988 & 102.790 & 86.31 & 85.98 & 0.29 & 0.844 & 0.428 & 32.0 \\ 
E2 & & & & 86.31 & 0.33 & & 0.811 & \\
E2 & & & & 86.71 & 0.34 & & 0.613 &  \\
E3 & 14.168 & 103.850 & 86.31 & 86.10 & 0.57 & 1.060 & 0.577 & 22.1 \\
E3 &  & &  & 86.35 & 0.25 &  & 0.663 &  \\
E3 &  & &  & 86.70 & 0.35 &  & 0.767 &  \\
\multicolumn{10}{l}{\bf Cloudlet 33}\\
E1 & -60.445 & 122.938 & 86.61 & 86.66 & 0.38 & 0.168 & 0.171 & 7.5 & -0.12$\pm$0.03  & 0.16$\pm$0.01  \\ 
E2 & -60.806 & 122.658 & 86.75 & 86.72 & 0.36 & 0.180 & 0.181 & 12.9 & (-1.8$\pm$0.4) & (2.5$\pm$0.1)\\ 
E3 & -61.147 & 123.733 & 86.75 & 86.76 & 0.31 & 0.242 & 0.239 & 14.7 \\
\multicolumn{10}{l}{\bf Cloudlet 34}\\
E1 & 11.305 & 117.839 & 87.67 & 87.64 & 0.25 & 0.433 & 0.428 & 11.5 & 0.11$\pm$0.03  & 0.12$\pm$0.01 \\
E2 & 11.709 & 117.754 & 87.62 & 87.65 & 0.33 & 0.152 & 0.150 & 7.8 & (1.7$\pm$0.4) & (1.8$\pm$0.1)\\
E3 & 11.855 & 118.634 & 87.72 & 87.66 & 0.33 & 0.220 & 0.213 & 15.8 \\
\end{longtable}

\begin{figure*}[h]
\centering
\includegraphics[scale=0.57, trim={0 3cm 0 2cm},clip]{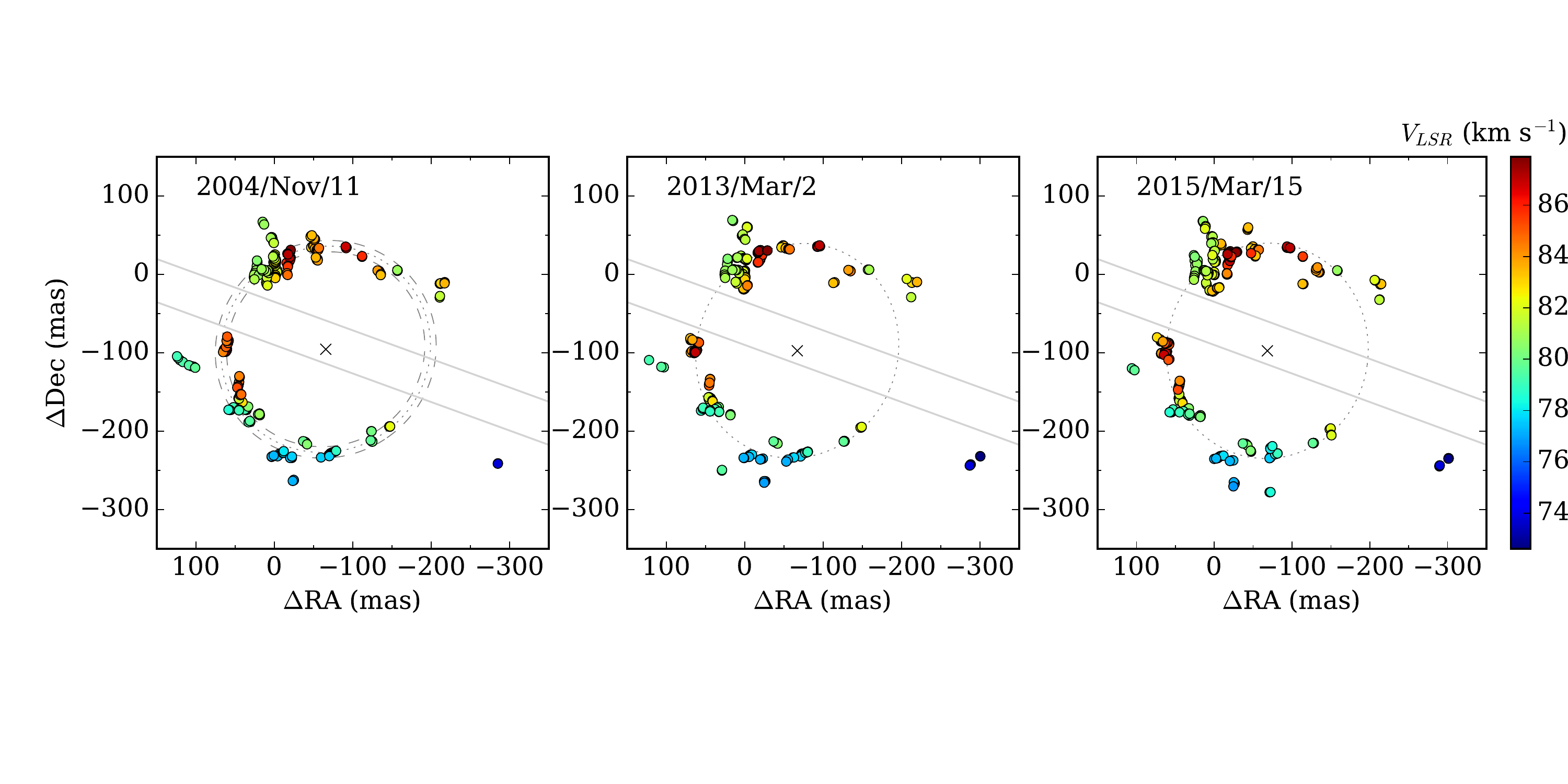}
\caption{Three epoch EVN maps of 6.7~GHz methanol maser spots detected in G23.657$-$00.127. The coordinates are relative to the brightest spots in each epoch (Table~\ref{table1}). The colours of circles relate to the LSR velocities as shown on the right bar. 
The dotted ellipses trace the best fits to all spots, except the SW blue-shifted ones, detected at each epoch (Table~\ref{ellipses}). The dashed ellipses indicate the width of the ring as derived in the first epoch data (Bartkiewicz et al.~\cite{b05}). The grey lines at  PA\,$\sim$80\degr
(N to E) delineate the emission-free part of the maser ring discussed in Sect.~4.2.}
\label{three}
\end{figure*}

\begin{figure}[h]
\centering
\includegraphics[width=0.49\textwidth]{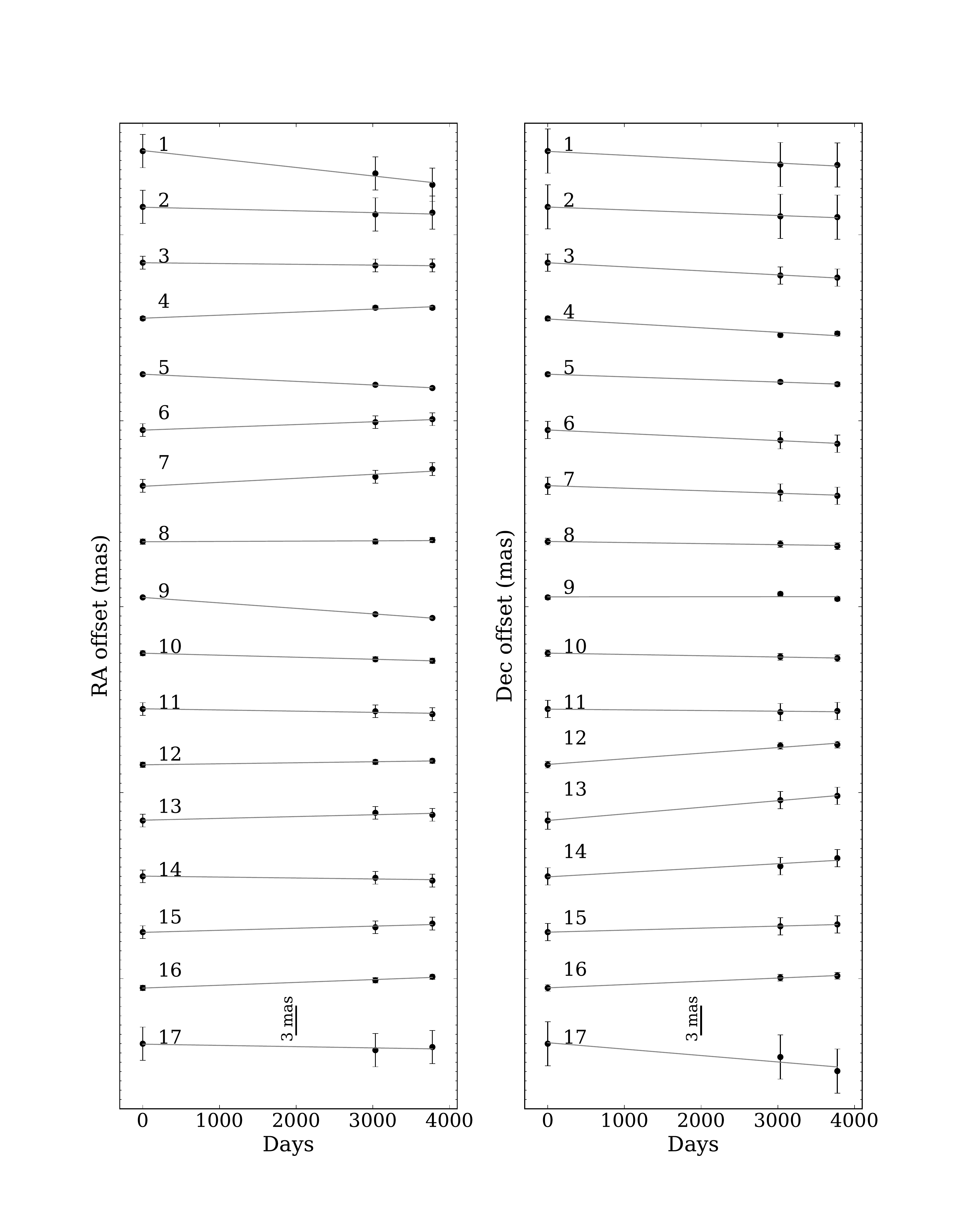}
\includegraphics[width=0.49\textwidth]{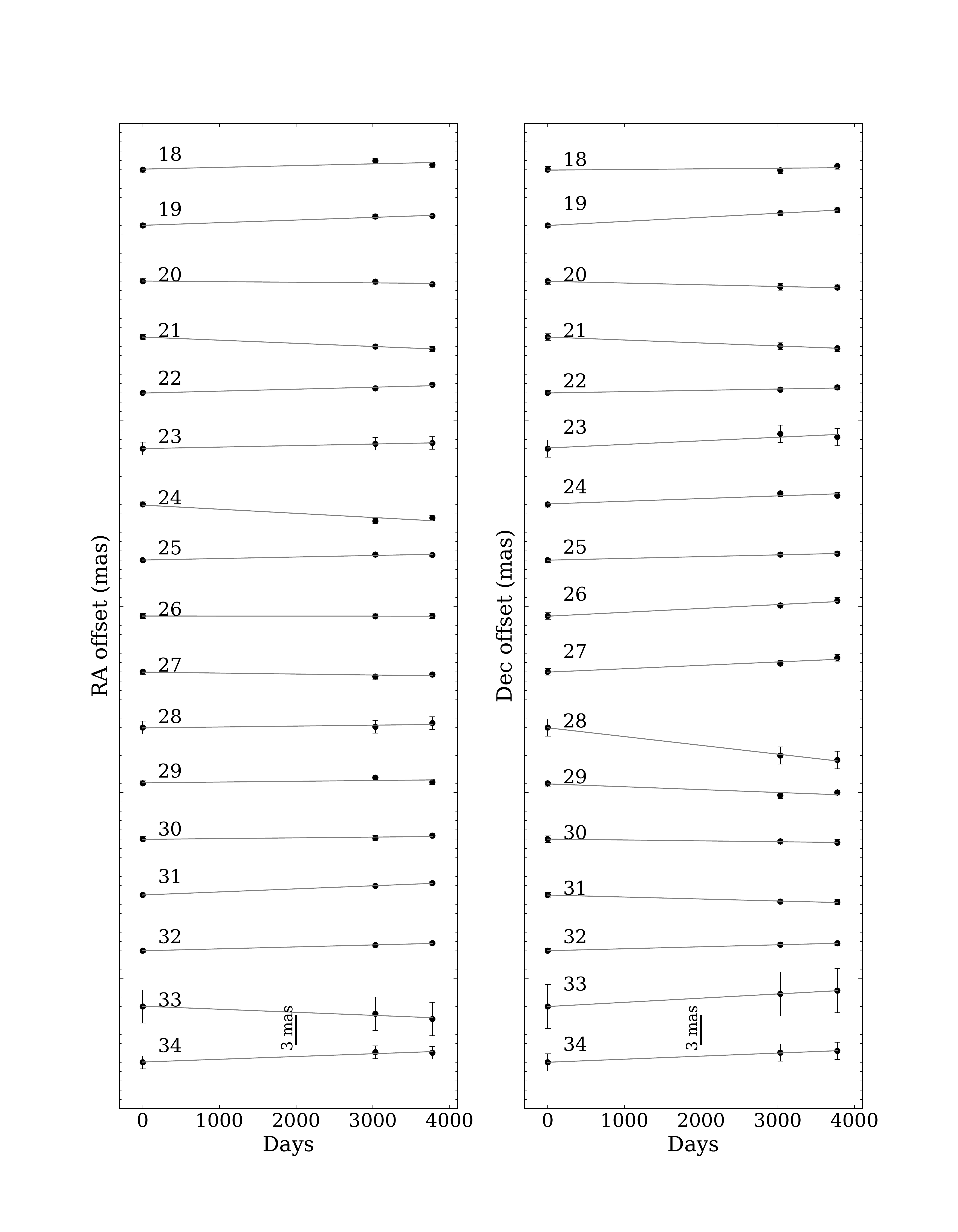}
\caption{Positional variations of the methanol maser cloudlets against time. The straight line shows the relative proper motion of cloudlet estimated by linear fitting. The cloudlets are labelled as in Table \ref{clouds}.
The error bars are magnified by a factor of 10 for clarity, and the time is relative to the first epoch.}
\label{linearmotions}
\end{figure}

\end{document}